# Plasma electron-hole kinematics: momentum conservation

I H Hutchinson and C T Zhou

May 20, 2016

**Abstract**

We analyse the kinematic properties of a plasma electron hole: a non-linear self-sustained localized positive electric potential perturbation, trapping electrons, that behaves as a coherent entity. When a hole accelerates or grows in depth, ion and electron plasma momentum is changed both within the hole and outside it, by an energization process we call jetting. We present a comprehensive analytic calculation of the momentum changes of an isolated general one-dimensional hole. The conservation of the total momentum gives the hole's kinematics, determining its velocity evolution. Our results explain many features of the behavior of hole speed observed in numerical simulations, including self-acceleration at formation, and hole pushing and trapping by ion streams.

## 1 Introduction

Electron holes are soliton-like localized electric potential maxima which self-consistently maintain themselves through a deficit of trapped electron phase-space density. They are frequently observed in space plasmas ([1, 2, 3, 4, 5, 6, 7, 8, 9, 10]), and in numerical simulations of kinetic electron instabilities ([11, 12, 13, 14, 15, 16, 17, 18, 19]). Under the influence of changes in the background plasma, such holes can change their speed or potential amplitude (depth). The resulting time dependence of the potential structure gives rise to changes in the electron and ion momentum through a process that is sometimes called "energization". In the present context we shall use the more relevant expression "jetting", since the momentum changes rather than the energy changes are our main concern. The purpose of this paper is to establish analytic expressions for the magnitude of the overall momentum changes for both ions and electrons. Since the electromagnetic momentum of an isolated electrostatic potential structure can be ignored, the total particle momentum must be conserved by the hole kinematics. The potential merely transfers momentum from one species to another, or one phase-space (position or velocity) region to another. Overall momentum conservation therefore enables us to deduce the kinematics of holes: their acceleration under some assumed potential evolution. We thereby explain the sometimes surprising hole behavior observed in numerical simulations.

Our treatment is entirely one-dimensional in space and velocity. Its applicability is therefore to holes that do not experience transverse instability or other breakup mechanisms[20, 21] on relevant timescales. That usually requires a strong magnetic field. Merging and splitting of one-dimensional holes is also sometimes observed but we do not address that possibility here, choosing to focus for clarity on a single isolated hole. Indeed, although there are well-motivated analytical mathematical forms for electron holes, for example that of Schamel[22, 23] sometimes called the Maxwell-Boltzmann hole, our analysis is almost entirely unaffected by the precise size or shape of the hole. We simply take a given potential of the hole to be characterized in amplitude by spatial potential integrals whose form our analysis discovers, and a propagation velocity. We consider this generality a strength of our approach.

Ground-breaking research on the topic of hole kinematics was undertaken by Dupree[24, 25]. However, his perspective was different; it was to explain *ion* holes, and to do so for simulations with extremely low artificial mass-ratio $m_i/m_e = 4$. Use of low mass-ratio was entirely understandable given the computing power limitations of the day. But it meant that there was no significant velocity scale separation between ions and electrons. Both velocity distributions were broad and overlapped the hole velocity. Physical mass ratios, by contrast, lead to ion velocity spreads that are far smaller than thermal electron velocities, and often much smaller than hole velocities. In practice therefore, electron holes and narrow velocity streams of ions are more characteristic. That is our perspective. It provides a substantially more transparent approach, avoiding the heavy algebra, early approximations, and questionable parts-integrations of Dupree. We reproduce some of his results, but we find essential leading-order terms that his treatment omitted.

We apply our results to interpret realistic mass-ratio simulations of electron holes. The things we are able to explain are (1) fast holes accelerating at the same rate as electrons in a background electric field[18]; (2) slower holes, without background electric field, able to be pushed (but not much pulled) by artificially-imposed slow ion speed changes[26]; (3) hole startup transients with electron deficits that suddenly accelerate deep holes to approximately the electron thermal speed[16, 27, 19]; (4) shallow hole startup with uniform density that accelerates the hole instead to about $c_s(m_i/m_e)^{1/4}$[26]; (5) hole velocity trapping between two ion streams[18].

We consider collisionless plasmas, in which the distribution function is constant along an orbit. The electron phase-space is illustrated in Fig. 1. The orbits in phase space are the contours of constant distribution function and, in steady state, also the contours of constant kinetic plus potential energy. A potential energy ($q\phi$) well is present, because the electrons have negative charge $q = -e$ and the electric potential $\phi$ is positive. It causes a region of trapped, finite, orbits on which the distribution function is determined during the formation process, rather than by connection to the distant plasma. Reduced electron distribution function value $f(v, x)$ on the trapped orbits is the primary cause of the hole's positive potential. But the ion density response is also important. Self-consistent electron holes satisfy Poisson's equation $d^2\phi/dx^2 = e(\int f dv - n_i)/\epsilon_0$, in which the electron distribution function is a function of total energy $\frac{1}{2}mv^2 + q\phi$. For details of an important example hole form see references [22, 23]. Holes are free to move and often have quite high speeds. This paper's intent is not to solve for holes' structure, but instead to understand more generally what determines their velocity and acceleration.

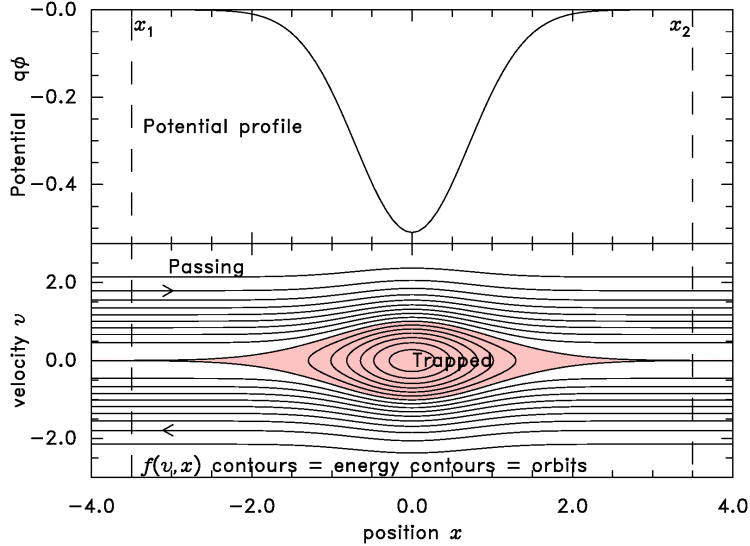

Figure 1: Schematic illustration of the form of a hole in electron phase space. The potential *energy* is plotted, the product of charge ($q = -e$) and electric potential. Electrons are the trapped species. The units of velocity are approximately electron thermal speed.

The structure of the paper is as follows. In section 2 we derive the momentum changes of an ion stream of specified velocity, arising from hole acceleration or growth. Section 3 analyzes the momentum changes of a broad distribution of velocities, which is essential for electrons. It also derives shallow-hole approximations useful when the hole potential is much less than the temperature. Section 4 shows how to use the balance between these momentum change rates to deduce the hole velocity evolution, and thereby explains several simulation observations.

## 2 Ion stream momentum changes

We consider the effects on an ion stream of acceleration or amplitude growth of a finite-extent potential structure $\varphi(x)$ that we call the hole. For notational convenience we take $\varphi$ to be the *potential energy* $q\phi$ of the species under consideration, whose charge is $q$ in the electric potential $\phi$. We denote by $x_1$ and $x_2$ the positions of the beginning and the end of the potential structure, at both of which $\varphi = 0$. Other quantities with subscripts 1 or 2 are naturally values at $x_1$ or $x_2$. It will become clear that the precise choice of $x_1$ and $x_2$ is unimportant provided the potential is uniform outside the interval $(x_1, x_2)$. However, it *is* important that the potential has the same value at $x_1$ and $x_2$. The presumption $\varphi_1 = \varphi_2$ constitutes our definition of, and limitation to, a "hole" and excludes what is normally called a "double layer" in which the potentials either side of the structure are different.

We consider two different frames of reference: the *hole* frame, and the *inertial* frame. The hole frame accelerates with the hole. The inertial frame does not, but the hole's instantaneous initial speed in it is zero. The ions are the untrapped species for an electron hole. Their phase-space, in the hole frame (like Fig. 1), is illustrated in Fig. 2. Some of the orbits are passing and some are reflected. We address one orbit at a time, corresponding to a particular distant velocity $v_1$. Velocities in the hole frame are denoted $v$; velocities in the inertial frame $v'$ where we need to be explicit. We note that $x_1$ and $x_2$ are constants in the *hole* frame, not in the *inertial* frame. The hole velocity in the inertial frame is $U$ and the acceleration is $\dot{U}$. Over dots denote time derivatives, and when applied to the local quantities $\varphi$ or $v$ mean partial derivative with respect to time at constant position in the hole frame.

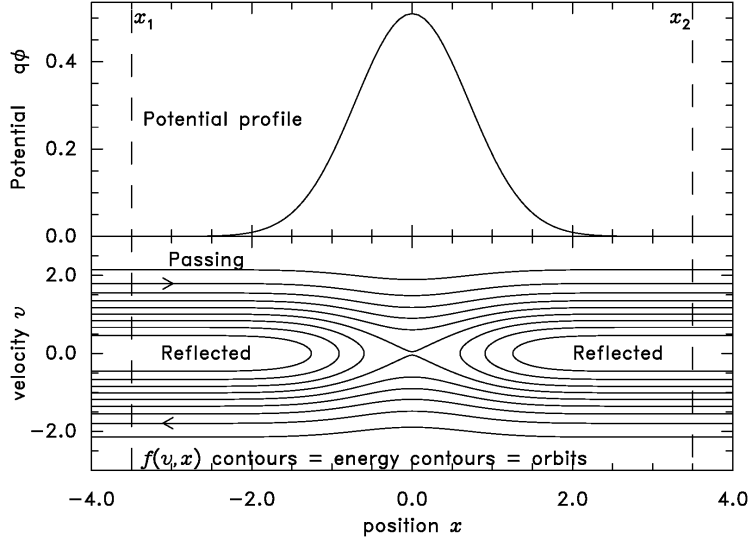

Figure 2: Schematic illustration of the ion phase space orbits in the rest frame of an electron hole. Potential *energy*, $\varphi = q\phi$, is plotted. Ions are the untrapped species in a positive electric potential because their charge is positive ($q = +e$). The units of velocity are approximately ion acoustic speed.

Our initial aim is to calculate the total rate of change of the momentum of the ion stream as a result of the presence, acceleration, and growth of the hole.

## 2.1 Ion jetting by hole acceleration

The transit time across the hole of an ion in the stream is

$$\delta t = t_2 - t_1 = \int_{x_1}^{x_2} \frac{dx}{v}. \tag{1}$$

We treat first only ions that are *passing*, not reflected by the potential. Reflected streams will be addressed later, in section 2.5. In positive (ion repelling) electric potentials, there are no trapped ions, though the passing-ion treatment is unaffected by the sign of $\varphi$.

In the accelerating hole frame, ions experience an extra reaction force so that

$$m\frac{dv}{dt} = -\frac{\partial \varphi}{\partial x} - m\dot{U}. \tag{2}$$

For constant $\dot{U}$ this can immediately be integrated to give the equivalent energy equation in the accelerating frame

$$\frac{1}{2}mv^2 + \varphi + m\dot{U}x = constant. \tag{3}$$

Consequently
$$v_2 - v_1 = -\dot{U}\frac{2}{v_2 + v_1}\int_{x_1}^{x_2} dx \tag{4}$$

The velocity difference *in the inertial frame* that occurs during an ion's transit across the hole is then
$$v'_2 - v'_1 = v_2 - v_1 + \delta t \dot{U} = \dot{U}\left(\frac{-2}{v_2+v_1}\int_{x_1}^{x_2} dx + \int_{x_1}^{x_2}\frac{dx}{v}\right). \tag{5}$$

If all ions that enter at $x_1$ could be considered also to exit at $x_2$, then the net (inertial frame) momentum outflow rate from the hole region would be simply
$$mn_1 v_1(v'_2 - v'_1) = mn_1\dot{U}\int_{x_1}^{x_2}\frac{-2v_1}{v_2+v_1} + \frac{v_1}{v}\, dx, \tag{6}$$

where $n$ denotes the stream density. However, because of the change in the ion stream velocity in the hole frame, there is additional ion accumulation within the hole. Therefore, although eventually all individual ions that enter do exit, over any finite duration more ions must enter than leave. For small durations, the number accumulated is proportional to duration, so an accumulation rate exists. Consequently there is an additional term that must be added to the momentum outflow rate to account for the ion accumulation. We now calculate it.

## 2.2 Ion accumulation by hole acceleration

In a static hole for which $\dot{U} = \dot{\varphi} = 0$ the flow-through of ions of density $n$ is governed by static continuity: $nv = n_1 v_1$, and so the number of ions in the hole is
$$N = \int_{x_1}^{x_2} n\, dx = n_1 v_1 \int_{x_1}^{x_2}\frac{dx}{v}. \tag{7}$$

When the hole is accelerating, rather than static, we can continue to use this expression as an approximation, only to the extent that the transit time of ions $\delta t$ is much shorter than the typical acceleration timescale $v/\dot{U}$. To do so will give us expressions correct only to lowest order in $\delta t \dot{U}/v$, the relative hole velocity change in a transit time. We are therefore considering changes that are slow compared to an ion transit time. To this same degree of approximation (lowest order in $\delta t \dot{U}/v$) one can also quickly show that $(v_2 + v_1)/2$ may be written as $v_1$, which can be used in eq. (6).

In that case, assuming $\dot{\varphi}$ is zero (in the hole frame),
$$\dot{N} = \frac{d}{dt}\int_{x_1}^{x_2} n dx = \frac{d}{dt}\left(n_1 v_1 \int_{x_1}^{x_2}\frac{dx}{v}\right) = n_1 \dot{v}_1 \int_{x_1}^{x_2}\left[\frac{1}{v} + v_1\frac{d}{dv_1}\left(\frac{1}{v}\right)\right] dx. \tag{8}$$

Here we have recognized that by our short-transit-time approximation $v$ is a function of $v_1$ and that is all the time dependence it has. Since $vdv = v_1 dv_1$, and $\dot{v}_1 = -\dot{U}$, we obtain
$$\dot{N} = -n_1\frac{\dot{U}}{v_1}\int_{x_1}^{x_2}\left[\frac{v_1}{v} - \left(\frac{v_1}{v}\right)^3\right] dx. \tag{9}$$

To account for the ion accumulation contribution to momentum outflow rate in the inertial frame, we must subtract from the expression of eq. (6) for $mn_1v_1(v_2' - v_1')$ the rate of momentum accumulation by ion accumulation. Since instantaneously $v_1' = v_1$, the additional term is $m\dot{N}v_1' = m\dot{N}v_1$. The combined result (approximating $v_2 + v_1$) is a net momentum outflow rate in the inertial frame

$$\dot{P}_o = mn_1\dot{U}\int_{x_1}^{x_2}\left[-1 + 2\frac{v_1}{v} - \left(\frac{v_1}{v}\right)^3\right]dx, \tag{10}$$

where $v(x) = \sqrt{v_1^2 - 2\varphi(x)/m}$, $v_1$ is (of course) independent of $x$, and $v$ and $v_1$ have the same sign.

## 2.3 Contained and total ion momentum

In addition to the rate of change of momentum in the outer region, i.e. the momentum outflow rate we've just calculated, the total must include the rate of change of the ion momentum contained in the hole.

To lowest order in the transit time $\delta t \dot{U}/v$, the continuity equation requires that the contained ion momentum *density*, $mnv$, in the hole frame remains uniform and equal to the external value $mn_1v_1$. There is, nevertheless, a change in the inertial frame contained momentum, because the hole is accelerating and its density is different from the external density. The total momentum in the inertial frame is equal to the momentum in the accelerating hole frame $\int_{x_1}^{x_2} mnvdx = mn_1v_1\int_{x_1}^{x_2}dx$ plus the mass $\int_{x_1}^{x_2} mndx$ times the hole velocity. Combining these effects and using $\dot{v}_1 = -\dot{U}$, $n/n_1 = v_1/v$, and $U = 0$ instantaneously, we obtain the rate of change of contained momentum in the inertial frame as

$$\dot{P}_c = m\frac{d}{dt}\int_{x_1}^{x_2}(nv + nU)dx = m\dot{U}\int_{x_1}^{x_2}(-n_1 + n)\,dx = mn_1\dot{U}\int_{x_1}^{x_2}\left(-1 + \frac{v_1}{v}\right)dx. \tag{11}$$

Adding the outflow and contained ion momentum rates of change we get a total

$$\dot{P}_o + \dot{P}_c = mn_1\dot{U}\int_{x_1}^{x_2}\left[-2 + 3\frac{v_1}{v} - \left(\frac{v_1}{v}\right)^3\right]dx, \tag{12}$$

in which we have made *no approximation in respect of the magnitude of $\varphi$*. All we have assumed is that there is no reflection and that the transit time is short. This is an absolutely critical difference from Dupree's treatment[25]. He made small-$\varphi$ approximations early on and embedded the consequent Taylor expansions into his derivation. That led him to expressions which have zero momentum transfer to the untrapped species other than by reflection. There is in fact *non-zero* momentum transfer to *passing* particles. However when $mv_1^2 \gg 2\varphi$, it is *second order* in the small quantity $\varepsilon \equiv 2\varphi/mv_1^2$, not first order. One can see this from the Taylor expansions $\frac{v_1}{v} = \frac{1}{(1-\varepsilon)^{1/2}} = 1 + \frac{1}{2}\varepsilon + \frac{3}{8}\varepsilon^2 + O(\varepsilon^3)$, and $\left(\frac{v_1}{v}\right)^3 = \frac{1}{(1-\varepsilon)^{3/2}} = 1 + \frac{3}{2}\varepsilon + \frac{15}{8}\varepsilon^2 + O(\varepsilon^3)$, which lead to

$$-2 + 3\frac{v_1}{v} - \left(\frac{v_1}{v}\right)^3 = -\frac{3}{4}\varepsilon^2 + O(\varepsilon^3) \simeq -3\left(\frac{\varphi}{mv_1^2}\right)^2, \tag{13}$$

to lowest (nonzero) order. The corresponding momentum rate of change approximation is

$$\dot{P}_o + \dot{P}_c \simeq -3mn_1\dot{U} \int_{x_1}^{x_2} \left(\frac{\varphi(x)}{mv_1^2}\right)^2 dx. \tag{14}$$

Summarizing thus far, eq. (12) or approximately (14) provides the total rate of change of momentum for a velocity stream that is unreflected, taking the transit time parameter $\delta t \dot{U}/v$ to be small. Clearly, the expression can be immediately integrated over a range of $v_1$ to obtain the total for a broad unreflected distribution function, but we defer that treatment till section 3.

## 2.4 Momentum change due to hole growth

Now instead of considering an accelerating hole we treat the case $\dot{U} = 0$, but $\dot{\varphi} \neq 0$, a growing stationary hole. In this second case the hole is stationary in the initial inertial frame and remains so. $U = 0$, so there is less distinction between $v$ and $v'$, recall though that $v$ is in the hole rest frame.

The energy increase of an ion transiting the hole from $x_1$ to $x_2$ is simply

$$\frac{1}{2}m(v_2^2 - v_1^2) = \int_{t_1}^{t_2} \dot{\varphi} dt = \int_{x_1}^{x_2} \dot{\varphi} \frac{dx}{v}. \tag{15}$$

Hence, to first order in transit time using the $v_2 + v_1$ approximation

$$v_2 - v_1 = \frac{1}{mv_1^2} \int_{x_1}^{x_2} \frac{v_1}{v} \dot{\varphi} \, dx. \tag{16}$$

We must also account for accumulation of ions as the hole depth grows.

$$\dot{N} = n_1 \int_{x_1}^{x_2} \frac{\partial}{\partial t}\left(\frac{v_1}{v}\right) dx = n_1 \int_{x_1}^{x_2} \dot{\varphi} \frac{d}{d\varphi}\left(\frac{v_1}{v}\right) dx = n_1 \frac{1}{mv_1^2} \int_{x_1}^{x_2} \left(\frac{v_1}{v}\right)^3 \dot{\varphi} \, dx. \tag{17}$$

The last equality of eq. (17) recognizes that $\frac{d}{d\varphi}(\frac{v_1}{v}) = \frac{v_1}{mv^3}$.

There is no change of internal momentum density ($\dot{P}_c$) from hole depth growth because $nv$ remains uniform. Therefore the two contributions to the momentum outflow give the total in this case

$$\begin{aligned}\dot{P}_g &= mn_1v_1(v_2 - v_1) - m\dot{N}v_1 \\ &= \frac{n_1}{v_1} \int_{x_1}^{x_2} \left[\left(\frac{v_1}{v}\right) - \left(\frac{v_1}{v}\right)^3\right] \dot{\varphi} \, dx \\ &= mn_1v_1 \int_{x_1}^{x_2} \left[\left(\frac{v_1}{v}\right) - \left(\frac{v_1}{v}\right)^3\right] \frac{\dot{\varphi}}{mv_1^2} \, dx.\end{aligned} \tag{18}$$

This is the momentum change due to hole growth without approximation in respect of $\varphi$. The integrand is clearly negative for positive $\varphi$ and $\dot{\varphi}$. For a shallow hole one might now consider a Taylor expansion in $\varepsilon$ and find

$$\left(\frac{v_1}{v}\right) - \left(\frac{v_1}{v}\right)^3 = -\varepsilon + O(\varepsilon^2). \tag{19}$$

Then to lowest order in $\varepsilon = 2\varphi/mv_1^2$

$$\dot{P}_g \approx -n_1 \frac{1}{mv_1^3} \int_{x_1}^{x_2} 2\varphi\dot{\varphi}\, dx = -mn_1 v_1 \int_{x_1}^{x_2} \frac{\partial}{\partial t}\left(\frac{\varphi}{mv_1^2}\right)^2 dx. \qquad (20)$$

## 2.5 Reflected ion stream momentum change

If an ion stream encounters a potential structure higher than its kinetic energy (in the hole frame), it is reflected. In that case, the momentum transfer rates of the prior sections are inapplicable. Instead, in the hole frame, the momentum change of each reflected ion is $-2mv_1$. The rate at which ions encounter the hole is $n_1|v_1|$. Therefore the total rate of ion momentum change caused by the hole is

$$\dot{P}_r = -2n_1|v_1|v_1. \qquad (21)$$

Notice that no acceleration or growth are required in order for this force to exist. Since by the presumption of the previous sections $\dot{U}\delta t$ and $\dot{\phi}\delta t$ are small quantities, the reflection force, if it exists, will predominate over all others. Consequently, unbalanced reflection can occur only transiently. Furthermore, even though an equilibrium can be constructed consisting of balanced reflection of particle streams from both sides of the hole, that equilibrium is *unstable*, as we will show in section 4.4.

# 3 Distributed electron and ion momentum changes

Our analysis in the previous section has given us the rate of change of momentum of a passing particle stream under the influence of an accelerating and growing isolated potential structure. Although we have called the particles 'ions' in that section, there is nothing about the analysis that is specific to the heavier or untrapped species. Therefore the analysis applies *equally well* to a passing electron stream.

An electron hole's electron momentum has two aspects that are not covered by the previous section's analysis. First, the passing electron distribution function cannot be represented by a single stream of electrons, because a hole never moves substantially faster than the electron thermal speed. By contrast, electron holes can and usually do move faster than the ion thermal speed, so there exist situations where a single ion stream gives a good approximation of the entire kinetic ion response. The second aspect not yet covered is that there are trapped electrons.

## 3.1 Background force field

Before addressing explicit electron jetting effects that arise when a hole moves in a fixed background electron distribution, we show that if ion influence can be ignored, electron holes naturally move in phase space along the background orbits of the electrons that constitute them. And we derive the ion momentum conservation term when ions experience background acceleration $\dot{v}_1$.

We have in mind a situation where there exists a weak background force field (e.g. electric field) that can be approximated as uniform in the vicinity of the hole. It causes an acceleration of the *background, external* distribution function, which is therefore changing in time. Weakness of the force field means the background change in a transit time is small. The equation of motion of all electrons is simply

$$m\frac{dv}{dt} = -\frac{\partial \varphi}{\partial x} + F. \tag{22}$$

where $F$ is the background force. We observe (c.f. eq. 2) that this is the equation of motion without a background force, expressed in a frame accelerating at $\dot{u} = -F/m$. Therefore if there exists a stationary hole solution with self-consistent orbits and potential in an inertial frame without a force field, then there exists a totally equivalent solution viewed in an accelerating frame, in which all the particles are acted on by the force $F = -\dot{u}m$. In that frame of reference the hole is accelerating at $\dot{U} = -\dot{u}$. Thus Galilean relativity shows that if every particle is acted on by a force $\dot{U}m$, then the hole (as well as the particles) accelerates at $\dot{U}$.

If the background force field is gravity, then ions as well as electrons satisfy this Galilean equivalence. If, however, the background force is electric, then ions do not satisfy this equivalence. Therefore an electron hole accelerates at the same rate as the background electrons *only when ions can be ignored*. This condition is satisfied only if the hole is moving relative to the ions faster than $\sim (m_i/m_e)^{1/4}$ times the sound speed, as we'll show in section 4.1. In that case, the result can be conceptualized alternatively as saying that the effective charge-to-mass ratio of an electron hole is equal to that of electrons.

Equivalence arguments like this also give us the ion momentum rate for a stationary hole in a background ion stream that is accelerating by a background force field $F_i$. In a frame accelerating with the background ions there is no net force on the ions. So they respond within this frame to the hole acceleration rate ($\dot{U} = -F_i/m$) equal to minus the ion acceleration rate, giving a net momentum outflow rate eq. (10). In the hole frame (which is not accelerating in this situation) the contained momentum density is uniform, although increasing with time because of the background acceleration, giving $\dot{P}_c = -mn_1\dot{U}\int_{x_1}^{x_2} dx$. The background force exactly balances the background acceleration. However, within the hole the density is increased relative to the background density, and so the applied force on the ions exceeds their momentum density rate of increase. The hole is responsible for absorbing this additional momentum input. Starting from setting the total electron plus ion momentum increase rate equal to the force momentum input $\dot{P} = F_i \int_{x_1}^{x_2} n dx$, we can eliminate the background force term, reducing momentum conservation to $\dot{P} = 0$, by combining the force term into an *effective* contained ion momentum

$$\dot{P}_c = -mn_1\dot{U}\int_{x_1}^{x_2} dx - F_i \int_{x_1}^{x_2} n dx = mn_1\dot{U}\int_{x_1}^{x_2}\left(\frac{v_1}{v} - 1\right) dx. \tag{23}$$

This is equal to the expression (11) for contained momentum rate when a hole accelerates in a stationary ion stream. The resulting combined ion momentum rate expression is therefore the same as (12). However the meaning is different, because the current meaning of $\dot{U}$ is the stationary hole's acceleration *in the accelerating ion frame*, not in the inertial frame. We thus ought to use $\dot{U} = -\dot{v}_b$, where $\dot{v}_b$ is the acceleration of the background ions in the inertial frame, and write unambiguously

$$\dot{P}_b = -mn_1\dot{v}_b \int_{x_1}^{x_2} \left[-2 + 3\frac{v_1}{v} - \left(\frac{v_1}{v}\right)^3\right] dx, \tag{24}$$

This background-force momentum rate is the additional effective momentum rate contribution caused by an external field accelerating the background particles (regardless of what the hole actually does). If we add to it the momentum rate from a hole acceleration in the inertial frame, eq. (12), the total will be zero only if $\dot{v}_b = \dot{U}$, that is if the hole accelerates at the rate of background ions.

## 3.2 Full velocity distributions

Now we consider the additional features of electron dynamics and full velocity distributions. It is, in principle, straightforward to extend the ion stream analysis to cover a full distribution of passing particles. All we need to do is to replace in the prior equations the density $n_1$ by the density element $f_1(v_1)dv_1$ and then integrate $dv_1$ over the passing particles. Strictly speaking, there exists a forbidden region just outside the phase-space separatrix where the approximation of short-transit-time breaks down. The contribution from that region to the momentum of the *trapped species* (electrons for an electron hole, or ions for an ion hole) turns out to be unimportant. Problems arise for the *untrapped species*, which we defer.

We can immediately write down the momentum change of the passing electrons due to hole acceleration plus background acceleration, or growth as

$$\dot{P}_o + \dot{P}_c + \dot{P}_b = m(\dot{U} - \dot{v}_b) \int_{x_1}^{x_2} \int \left[-2 + 3\frac{v_1}{v} - \left(\frac{v_1}{v}\right)^3\right] f_1(v_1) dv_1 dx, \tag{25}$$

and

$$\dot{P}_g = m \int_{x_1}^{x_2} \int \left[\left(\frac{v_1}{v}\right) - \left(\frac{v_1}{v}\right)^3\right] \frac{\dot{\varphi}}{mv_1^2} f_1(v_1)v_1 dv_1 dx. \tag{26}$$

Since we have adopted the notation that $\varphi$ is the *potential energy* (charge times electric potential), the prior equations apply to either species without change of anything except mass (and background acceleration perhaps).

The momentum change of the trapped electrons (having speed smaller than the phase-space separatrix speed $v_s = \sqrt{-2\varphi/m}$) is even more straightforward. It arises simply from the acceleration of the trapped density, $\int_{-v_s}^{v_s} f_t(v)dv$, integrated over the hole spatial extent, less the background force on these particles

$$\dot{P}_t = m(\dot{U} - \dot{v}_b) \int_{x_1}^{x_2} \int_{-v_s}^{v_s} f_t(v)dv\, dx. \tag{27}$$

There is no change of trapped momentum due to hole growth, in the instantaneous rest inertial frame.

Reflected particles must also be included as a possibility. Reflected electrons will occur for an ion hole (negative electric potential, positive electron potential energy $\varphi$), and reflected ions from an electron hole. The momentum rate of change they experience is the velocity integral of the product of their rate of encountering the hole $f_1|v_1|$ and their individual momentum change $-2mv_1$

$$\dot P_r = -m2 \int_{-v_p}^{v_p} f_1(v_1)|v_1|v_1 dv_1, \tag{28}$$

where $v_p = \sqrt{2\psi/m}$ denotes the maximum reflected speed, determined by the maximum potential $\psi = \max(\varphi)$. We can also regard $v_p$ as the maximum separatrix speed, which occurs at the hole edge, $\varphi = 0$ for the untrapped species. The net reflection momentum rate is zero for symmetric $f_1$. It arises from two partial pressures $\int_0^{\pm v_p} f_1 v_1^2 dv_1$ acting on either side of the hole. So only the asymmetric part of the pressure matters to the net force.

The total electron momentum rate of change, referring to eqs. (25), (26), (27), and (28) is

$$\dot P = \dot P_o + \dot P_c + \dot P_b + \dot P_g + \dot P_t + \dot P_r. \tag{29}$$

An identical equation can also be considered to hold for ions using ion parameters, and the sum of the electron and ion totals is to be set to zero. An electron hole has non-zero ion reflection contribution $\dot P_r$ if the ion distribution extends to zero $v$. An ion hole has a non-zero trapped ion contribution $\dot P_t$. Velocity limits $v_s$ or $v_p$ that involve square-roots of negative values must be considered zero, eliminating the corresponding $\dot P$ term. In other words, an electron hole has only a trapped electron term, no reflected electron term, but it may have a reflected ion term, but no trapped ion term.

Although we *have* ignored the inadequacy of the short-transit-time approximation at small velocity, we have *not* made a shallow-hole approximation nor presumed $\varphi/mv^2$ to be small.

## 3.3 Trapped density deduced from charge balance

In order to evaluate the electron momentum, we need an expression for $f_t$ which determines via eq. (27) the $\dot P_t$ term. But actually that equation uses only the total number of trapped electrons. The total charge within a hole must be zero to make the potential gradient it generates at $x_1$ and $x_2$ zero. This fact constrains the trapped electron number because the *untrapped* electron (and ion) density can be calculated from the distribution function using the short-transit time approximation as

$$N = \int_{x_1}^{x_2} \int_{-\infty}^{\infty} \frac{v_1}{v} f_1(v_1) dv_1. \tag{30}$$

Thus hole total charge neutrality requires

$$0 = N_e - N_i = \int_{x_1}^{x_2} \left[ \int_{-v_s}^{v_s} f_t(v) dv + \int_{-\infty}^{\infty} \frac{v_1}{v} f_1(v_1) dv_1 - n_i \right] dx \tag{31}$$

We shall see later that the relative speed of the ions and the hole is generally sufficiently large that the ion (charge) density perturbation is small. We therefore ignore it to lowest order, and take $n_i = n_{e1} = \int_{-\infty}^{\infty} f_1(v_1) dv_1$, so that

$$\int_{x_1}^{x_2} \int_{-v_s}^{v_s} f_t(v) dv\, dx = \int_{x_1}^{x_2} \int_{-\infty}^{\infty} \left(1 - \frac{v_1}{v}\right) f_1(v_1) dv_1\, dx. \tag{32}$$

On this basis, using eq. (25), we can write the total electron momentum change due to acceleration as

$$\dot{P}_{eo} + \dot{P}_{ec} + \dot{P}_{eb} + \dot{P}_{et} = m(\dot{U} - \dot{v}_b) \int_{x_1}^{x_2} \int_{-\infty}^{\infty} \left[-1 + 2\frac{v_1}{v} - \left(\frac{v_1}{v}\right)^3\right] f_1(v_1) dv_1 dx. \quad (33)$$

This velocity integral can be expressed in closed form for a Maxwellian distribution $f_1(v_1) = \frac{n_1}{v_t\sqrt{\pi}} \exp(-v_1^2/v_t^2)$, where $v_t = \sqrt{2T_e/m_e}$. (It is reasonable to take this distribution to be unshifted even though we ignore ion density perturbation provided $c_s \ll U \ll v_t$.) Writing $\chi^2 = |\varphi|/T_e = v_s^2/v_t^2$ and $\zeta = v_1/v_t$, we find

$$\dot{P}_{eo} + \dot{P}_{ec} + \dot{P}_{eb} + \dot{P}_{et} = -m(\dot{U} - \dot{v}_b)n_1 \int_{x_1}^{x_2} h(\chi) dx \quad (34)$$

where the function $h$ is

$$\begin{aligned} h(\chi) &= -\frac{2}{\sqrt{\pi}} \int_0^{\infty} \left[-1 + \frac{2\zeta}{(\chi^2 + \zeta^2)^{1/2}} - \frac{\zeta^3}{(\chi^2 + \zeta^2)^{3/2}}\right] e^{-\zeta^2} d\zeta \\ &= -\frac{2}{\sqrt{\pi}}\chi + \left[(2\chi^2 - 1)e^{\chi^2} \text{erfc}(\chi) + 1\right]. \end{aligned} \quad (35)$$

The limiting behavior of $h(\chi)$ is $h(\chi) \to \chi^2 - \frac{2}{\sqrt{\pi}}\frac{4}{3}\chi^3$ as $\chi \to 0$, and $h(\chi) \to 1 - \frac{2}{\sqrt{\pi}}\frac{1}{\chi}$ as $\chi \to \infty$.

## 3.4 Shallow hole approximation, trapped species (electrons)

We now show rigorously how to obtain approximate expressions for the momentum time derivatives when the hole is shallow, meaning $\frac{2\varphi}{mv_t^2} \ll 1$, where $v_t$ denotes a typical velocity of the background distribution function, but we do not assume $f_1(v_1)$ to be Maxwellian. To avoid continually writing the $x$-integration, we deal with quantities that must be integrated over the hole spatial extent to give total momentum. They can be considered the momentum *density* rates denoted by lower case $\dot{p}$, and are functions of $x$, but it is really the total momentum, integrated over the hole, that has physical significance. The space integral should be considered implied. The rate attributable to acceleration, and background force, is

$$\dot{p}_o + \dot{p}_c + \dot{p}_b = m(\dot{U} - \dot{v}_b) \int \left[-2 + 3\frac{v_1}{v} - \left(\frac{v_1}{v}\right)^3\right] f_1(v_1) dv_1, \quad (36)$$

and the rate caused by hole growth is

$$\dot{p}_g = \dot{\varphi} \int \left[\left(\frac{v_1}{v}\right) - \left(\frac{v_1}{v}\right)^3\right] \frac{1}{v_1^2} f_1(v_1) v_1 dv_1. \quad (37)$$

We recall that $v_1$ and $v$ have the same sign, therefore if we replace $v$ with $+\sqrt{v_1^2 - 2\varphi/m}$ then we must replace $v_1$ in the fractions' numerators with $|v_1|$. (Alternatively, replace every $v$ with $|v|$.)

We deal first with the trapped species, so $\varphi$ is negative. Then $v_1/v \to 0$ as $v_1 \to 0$ and no singularity occurs for an integration across $v_1 = 0$. Notice that both the square bracket factors tend to zero as $v_1 \to \infty$ at constant $\varphi$. Those factors thus restrict the effective velocity range of the integrals to a few times $\sqrt{2\varphi/m}$, which is by presumption small compared with the velocity-width of $f_1$. Therefore we can approximate the integrals by Taylor expanding $f_1(v_1)$ (not the $v$ expressions) about $v_1 = 0$ as $f_1(v_1) \simeq f_0 + f'_0 v_1$. The two $f$-terms represent the locally even and odd parity parts of $f_1$, and must respectively be combined with the even or odd part of the rest of the integrand, discarding combinations which are overall odd, because they integrate to zero. The rest of the integrand is even for eq. (36) and odd for eq. (37). To account for the $|v_1|$ signs we then simply use $2\int_0^\infty dv_1$ for the overall even expression. Defining $\xi = v_1/\sqrt{\frac{-2\varphi}{m}}$, and using $v_s = \sqrt{\frac{-2\varphi}{m}}$ for the separatix speed, we find

$$\dot{p}_o + \dot{p}_c + \dot{p}_b \simeq 2f_0 v_s m(\dot{U} - \dot{v}_b) \int_0^\infty \left[-2 + \frac{3\xi}{(1+\xi^2)^{1/2}} - \frac{\xi^3}{(1+\xi^2)^{3/2}}\right] d\xi$$
$$= 2f_0 v_s m(\dot{U} - \dot{v}_b) \left[-2\xi + 3\sqrt{1+\xi^2} - \sqrt{1+\xi^2} - \frac{1}{\sqrt{1+\xi^2}}\right]_0^\infty. \quad (38)$$

Since the value at the upper limit ($\xi \to \infty$) is zero, this is simply

$$\dot{p}_o + \dot{p}_c + \dot{p}_b \simeq -2f_0 v_s m(\dot{U} - \dot{v}_b). \quad (39)$$

Similarly $\dot{p}_g$ requires only the odd part of $f_1$, namely $f'_0 v_1$, and gives

$$\dot{p}_g = 2f'_0 v_s \dot{\varphi} \int \left[\frac{\xi}{(1+\xi^2)^{1/2}} - \frac{\xi^3}{(1+\xi^2)^{3/2}}\right] d\xi = -2f'_0 v_s \dot{\varphi}. \quad (40)$$

If we add to these the trapped particle momentum rate density $\dot{p}_t = m\dot{U} \int_{-v_s}^{v_s} f_t(v) dv$, then the total rate of change of momentum density of the trapped species is

$$\dot{p}_o + \dot{p}_c + \dot{p}_b + \dot{p}_t + \dot{p}_g = m(\dot{U} - \dot{v}_b) \int_{-v_s}^{v_s} (f_t(v) - f_0) dv - 2f'_0 v_s \dot{\varphi}. \quad (41)$$

This agrees with equation (178) of Dupree [25]. His $M$ is our $m \int_{-v_s}^{v_s} (f_t(v) - f_0) dv$ integrated over space, his $nf'_0$ is our $f'_0$, and his $\gamma q \phi$ our $\dot{\varphi}$.

## 3.5 Shallow hole, untrapped species (ions)

At first glance, one might suppose that we can proceed in the same way for the untrapped species. In this case $\varphi$ is positive and for passing particles we must include only $|v_1| > v_p = \sqrt{2\varphi/m}$. Writing $\eta \equiv v_1/v_p$, one would then again require only the even parity part of $f_1$, ($f_0 = f_1(v_p)$) and obtain

$$\dot{p}_o + \dot{p}_c + \dot{p}_b = 2m(\dot{U} - \dot{v}_b) f_0 v_p \int_1^\infty \left[-2 + 3\frac{\eta}{\sqrt{\eta^2 - 1}} - \frac{\eta^3}{(\eta^2 - 1)^{3/2}}\right] d\eta. \quad (42)$$

However, the third term of the integrand gives a divergent integral at $\eta \to 1$. So this approach is incorrect. What it ignores is that close to reflection (near $\eta = 1$) the approximation of short transit time breaks down. As particle speed $v_1$ tends to $v_p$, becoming almost reflected, the transit time $\delta t$ increases without bound.

In any case, if $f_1(\pm v_p) \neq 0$ (which causes the the integral to diverge) then there are certainly reflected particles too. As we saw before, reflected particles give force larger than passing particles by one (inverse) power of $\dot{U}$. The reflection force does not depend on hole acceleration at all, only distribution asymmetry. Therefore, to lowest order, when reflected particles exist, the untrapped species passing particle momentum should be ignored. This was Dupree's approximation. If, by contrast, there are no reflected untrapped species, the higher order calculation must be pursued, and will give a convergent integral because $f_1(\pm v_p) = f_0 = 0$. In fact, we can deduce more: that it is impossible for a hole to satisfy the short-transit time approximation if there are (unbalanced) reflected particles. The very existence of a steady electron hole requires it to have speed relative to the ions sufficient to avoid ion reflection, practically regardless of its depth.

Eq. (36) can validly be applied to the untrapped species only when $f_0 = 0$, in which case eq. (42) is zero. Thus the small-hole approximation (42) is inapplicable. The only consistent small-hole approximation for untrapped species is instead to suppose that for all $v_1$ for which $f_1$ is non-zero, $v_p^2/v_1^2$ is small. In that case, one can instead expand $v$ in eq. (36) in the manner of section 2.3 and get the extension of eq. (14)

$$\dot{p}_o + \dot{p}_c + \dot{p}_b \simeq -3m(\dot{U} - \dot{v}_b) \int \left(\frac{\varphi}{mv_1^2}\right)^2 f_1(v_1) dv_1. \tag{43}$$

Similarly for hole growth we have the extension of eq. (20)

$$\dot{p}_g \simeq -m \int v_1 \frac{\partial}{\partial t} \left(\frac{\varphi}{mv_1^2}\right)^2 f_1(v_1) dv_1 \tag{44}$$

and the total is

$$\begin{aligned} \dot{p}_o + \dot{p}_c + \dot{p}_b + \dot{p}_g &\simeq -m \int \left(3(\dot{U} - \dot{v}_b) + 2\frac{v_1 \dot{\varphi}}{\varphi}\right) \left(\frac{\varphi}{mv_1^2}\right)^2 f_1(v_1) dv_1 \\ &= mn \left(\frac{\varphi}{m}\right)^2 \left[3(\dot{U} - \dot{v}_b) \left\langle \frac{1}{v_1^4} \right\rangle + 2\frac{\dot{\varphi}}{\varphi} \left\langle \frac{1}{v_1^3} \right\rangle \right]. \end{aligned} \tag{45}$$

If the distribution is narrow so that $f_1(v_1)$ can be considered approximately a delta function, the averaging signs $\langle \rangle$ can be removed and in the absence of a background force ($v_b = 0$), we can write $\dot{U} = -\dot{v}_1$. Then the ion momentum rate is zero and acceleration balances hole growth, when

$$\frac{3}{v_1}\frac{dv_1}{dt} = \frac{2}{\varphi}\frac{d\varphi}{dt}, \tag{46}$$

whose solution for the time dependence of $v_1$ with initial conditions labelled 0 is

$$\frac{v_1}{v_0} = \left(\frac{\varphi}{\varphi_0}\right)^{2/3}. \tag{47}$$

In words, a growing (or shrinking) shallow electron hole will cause zero ion momentum change when the ion velocity in the hole frame grows (or shrinks) proportional to the hole's potential depth to the two thirds power.

# 4 Deducing hole speed from momentum balance

## 4.1 Acceleration ion and electron momentum

At sound-speed velocity, ion momentum density exceeds thermal electron momentum density by a large factor $\sim \sqrt{\frac{m_i}{m_e}}$. On this basis one might suppose that electron momentum rate can be neglected in comparison with ions and so the hole must move in such a way as to zero the ion momentum rate of change. That's approximately true at low enough $v_1$. However, since the acceleration ion momentum term $\dot p_o + \dot p_c$ is proportional to $v_1^{-4}$, at an ion speed not very much higher than the sound speed the ion momentum rate will become smaller than the electron rate. The ion and electron rates due to hole acceleration are in the same direction; therefore the speed will scale more slowly than eq. (47) when electron momentum rate becomes significant.

Let us find the $v_1$ at which the ion and electron momentum rates are equal; that is the ion stream speed (in hole frame) $v_1$ at which the electron momentum rate due to *hole acceleration*

$$\dot p_{oe} + \dot p_{ce} + \dot p_{eb} + \dot p_t = -n_e m_e (\dot U - \dot v_b) h(\sqrt{|\varphi_e|/T_e}), \tag{48}$$

(where $h(\chi)$ is given in eq. (35)) is equal to the ion rate

$$\dot p_{oi} + \dot p_{ci} + \dot p_{ib} = -3 n_i \left(\frac{\varphi_i}{m_i v_1^2}\right)^2 m_i (\dot U - \dot v_b) = -3 n_i \left(\frac{c_s}{v_1}\right)^4 \left(\frac{\varphi_i}{T_e}\right)^2 m_i (\dot U - \dot v_b). \tag{49}$$

For singly charged ions (i.e. external electron and ion densities equal) and identical electron and ion background acceleration $\dot v_b$, equality leads immediately to

$$\frac{|v_1|}{c_s} = \left(\frac{m_i}{m_e}\right)^{1/4} \left(\frac{3}{h}\right)^{1/4} \left(\frac{|\varphi|}{T_e}\right)^{1/2} \equiv M_{ie}. \tag{50}$$

For hydrogen mass ratio, the predominant factor and the approximate value of $M_{ie}$ is $\left(\frac{m_i}{m_e}\right)^{1/4} = (1836)^{1/4} = 6.5$.

The additional factors in the normalized equality speed $M_{ie}$ become $(3|\varphi|/T_e)^{1/4}$ for shallow holes, since $h(\chi) \to \chi^2 = |\varphi|/T_e$, and become $\sim 3^{1/4}(|\varphi|/T_e)^{1/2}$ for deep holes, since $h(\chi) \to 1$. Strictly speaking, we must use spatially integrated quantities; so for shallow holes $h \to \int_{x_1}^{x_2} |\varphi|/T_e dx$, while $|\varphi_i|^2 \to \int_{x_1}^{x_2} |\varphi_i|^2 dx$. Then the spatial integration effects are represented correctly in the expression for $M_{ie}$ if we take $|\varphi|^{1/4} = [\int_{x_1}^{x_2} \varphi_i^2 dx / \int_{x_1}^{x_2} \varphi_i dx]^{1/4}$ (for shallow holes). The extra factors are fairly close to one unless the hole is of extreme depth or shallowness. For example, if $|\varphi|/T_e = 0.1$, then $h = 0.064$ and $M_{ie} \approx 5.4$; while a very deep hole $|\varphi|/T_e = 10$, gives $h = 0.67$, and $M_{ie} \approx 30$. In summary, shallow electron holes moving relative to ions faster than a few $c_s$ have electron momentum rate due to acceleration ($\dot U$) greater than ion rate.

If ions are artificially accelerated at rate $\dot v_b$ by a background force and electrons are not (so $\dot v_b = 0$ in the electron equations) we must set equal to zero the combined momentum rate. Recognizing $\dot U - \dot v_b = -\dot v_1$, we can write $\dot p = 0$ as

$$\frac{\dot U}{c_s} = \left(\frac{M_{ie}^4 c_s^4}{v_1^4}\right) \frac{\dot v_1}{c_s}, \tag{51}$$

which integrates for initial and final conditions $A$ and $B$ to

$$\left[\frac{U}{c_s}\right]_A^B = \left[\frac{M_{ie}^4 c_s^3}{-3v_1^3}\right]_A^B. \tag{52}$$

This result shows that although background acceleration of the ions (changing $v_1$ through $\dot{v}_b$) can cause the hole velocity $U$ to increase or decrease, there is strong asymmetry in the influence on $U$. For brevity, we call decreases in $|v_1|$ (moving the ion velocity towards the hole velocity) "pushing" and increases in $|v_1|$ (moving away) "pulling"; though one must recall this is referring to velocity, not position. Pushing can give an upper limit term $M_{ie}^4 c_s^3/3v_{1B}^3$ in eq. (52) that increases without bound. On this basis, the hole can be pushed to as large a $|U|$ as desired. However, when pulling, the upper limit term (at $B$) tends to zero as $|v_1| \to \infty$. Consequently, the maximum change of $U$ that can be obtained by pulling is given by the lower limit term representing the initial condition $M_{ie}^4 c_s^3/3v_{1A}^3$. For example, if the initial speed is $|v_1|/c_s = M_{ie}$, the maximum pulling $U$-change is $|\Delta U|/c_s = M_{ie}/3$ and if $|\varphi|/T_e = 0.1$, then $|\Delta U|/c_s \simeq 2$. If the initial $v_1$ is instead $2M_{ie}c_s$, then $|\Delta U|/c_s = M_{ie}/24$, which is practically negligible. Holes with initial $|v_1|$ significantly greater than $M_{ie}c_s$ can be pushed, but they can't significantly be pulled. However, pushing and pulling is reversible. If a hole is first pushed, somewhat reducing $|v_1|$, it then can be pulled back by an equal amount to the original $U$ value.

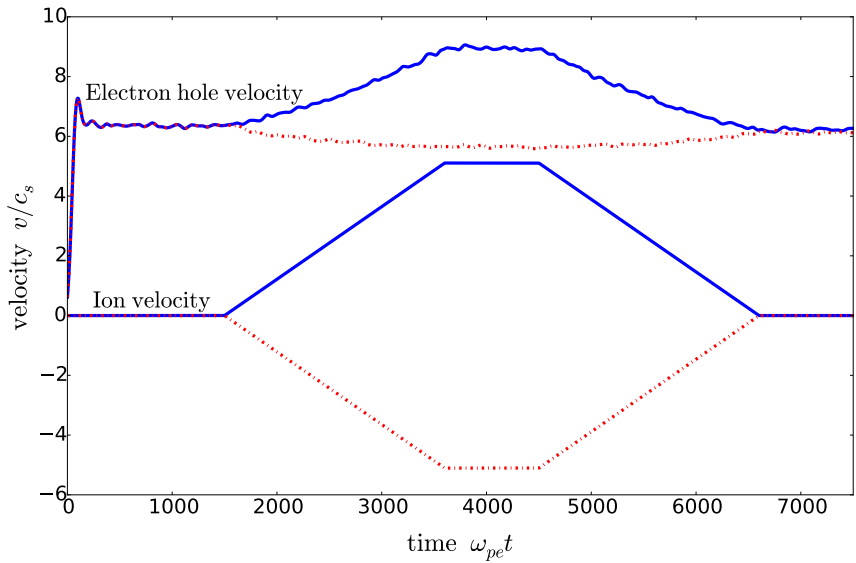

Figure 3: One-dimensional simulations showing asymmetry of hole velocity change by artificial ion acceleration. Hole depth $|\varphi_0|/T_e = 0.1$. Solid lines "pushing", dotted lines "pulling". [26]

These phenomena have been explored using PIC simulations[26] and the asymmetry and reversibility have been clearly demonstrated. Fig. 3 illustrates the results. Pulling gives only very minor hole acceleration, pushing gives much greater hole response, but both are reversible.

## 4.2 Hole-growth causing hole acceleration

When a hole grows in depth, the momentum rate due to hole-growth ion jetting, eq. (20) balances electron and ion acceleration jetting. The hole-growth electron jetting term is smaller and can be ignored, provided we stay near the peak of the electron distribution where $f'_0 \simeq 0$. We take $\dot{v}_b = 0$ in this section. The momentum balance in the rest frame of the background ions (so $U = -v_1$) is then

$$0 = -hm_e\dot{U} - 3\left(\frac{\varphi_i}{m_i v_1^2}\right)^2 m_i \dot{U} - \left(\frac{2\dot{\varphi}_i \varphi_i}{m_i v_1^3}\right). \tag{53}$$

Representing the hole depth as the ion passing speed (separatrix speed at infinity) $\sqrt{2|\varphi|/m_i} = v_p = \chi\sqrt{2}c_s$, the momentum balance then becomes

$$0 = -\frac{m_e}{m_i}h(\chi)\dot{U} - \frac{3}{4}\frac{v_p^4}{U^4}\dot{U} + \frac{v_p^3}{U^3}\dot{v}_p. \tag{54}$$

For a shallow hole, $h \approx \chi^2 = v_p^2/2c_s^2$. So

$$0 = -\frac{1}{2}\frac{m_e}{m_i}\frac{v_p^2}{c_s^2}\dot{U} - \frac{3}{4}\frac{v_p^4}{U^4}\dot{U} + \frac{v_p^3}{U^3}\dot{v}_p. \tag{55}$$

A particular solution (passing through the origin) to this differential equation relating $U$ and $v_p$ is

$$U^4 = \frac{5}{2}\frac{m_i}{m_e}c_s^2 v_p^2 = \frac{5}{2}\frac{m_i}{m_e}c_s^4\left(\frac{2|\varphi|}{T_e}\right), \tag{56}$$

that is, a parabolic relationship: $U^2 \propto v_p$.

Deep holes, by contrast, have $h \approx 1$ leading to a differential equation

$$0 = -\frac{m_e}{m_i}\dot{U} - \frac{3}{4}\frac{v_p^4}{U^4}\dot{U} + \frac{v_p^3}{U^3}\dot{v}_p, \tag{57}$$

which has a particular solution

$$U = \left(\frac{m_i}{4m_e}\right)^{1/4} v_p, \tag{58}$$

The entire family of solutions for different initial $U$, calculated numerically, using the full expression for $h$, is illustrated in Fig. 4. (The approximation of ignoring electron jetting due to hole growth doesn't really apply for $U \sim \sqrt{m_i/m_e}$; so caution is needed at high $U$.)

The effects of spatial integration of the electron and ion terms mean we should take (for shallow holes) $|\varphi| = [\int_{x_1}^{x_2} \varphi_i^2 dx/\int_{x_1}^{x_2} \varphi_i dx]$. For example if the potential shape is a Gaussian in space of maximum height $\varphi_0$. Then we should use $\int_{x_1}^{x_2} \varphi_i^2 dx/\int_{x_1}^{x_2} \varphi_i dx = \varphi_0/\sqrt{2}$ in place of $|\varphi|$ (e.g. in eq. (56)), and we can regard $v_p = \sqrt{\varphi_0/m_i}$, for purposes of calculating $U$.

Holes that remain shallow, $v_p \lesssim 1$, grown from a slow initial speed $U_0$ are accelerated only to $U \sim (m_i/m_e)^{1/4}c_s$. A shallow hole can nevertheless start with speed much larger than $(m_i/m_e)^{1/4}c_s$. Then the electron momentum is predominant; hole growth is unimportant; the condition for the applicability of section 3.1 is satisfied; and the hole accelerates in a background field at the rate of the external electrons. A hole that starts shallow (with $v_p$ small) at an intermediate initial speed $U_0 \gtrsim (m_i/m_e)^{1/4}c_s$, accelerates substantially only if it becomes deep ($v_p > 1$).

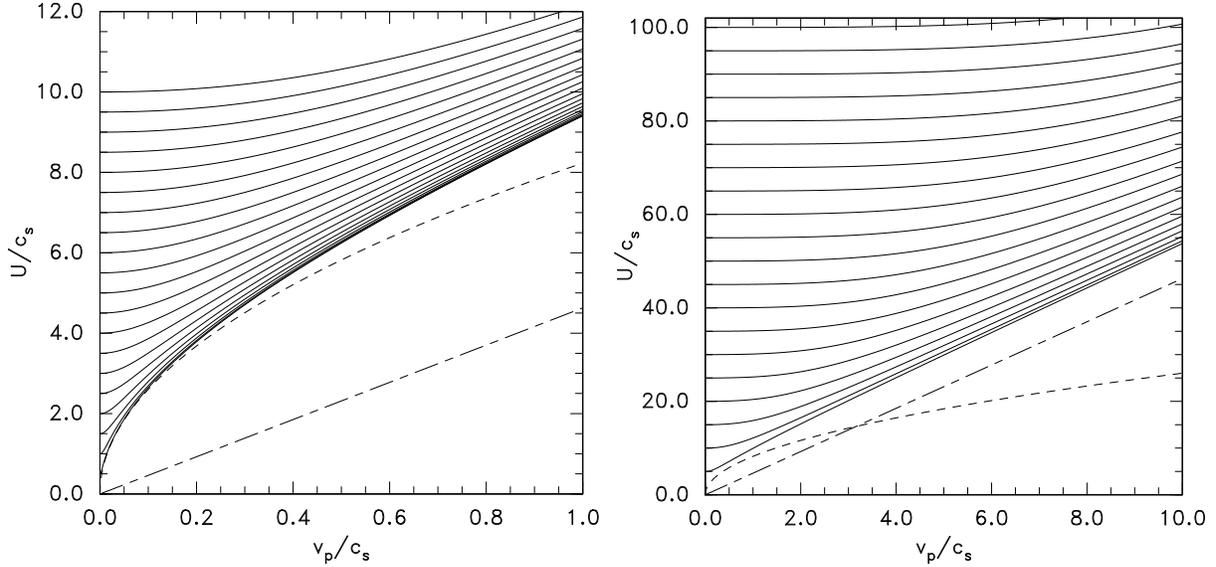

Figure 4: Growth curves of hole speed $U$ versus hole depth $v_p$ solving eq. (54) with $m_i/m_e = 1836$. The analytic approximation solutions, for shallow and deep holes, eqs. (56, 58), are shown with the dashed and dash-dot lines. All curves lie above them.

## 4.3 Initial transient hole acceleration

We now use the momentum expressions we've derived to estimate the self-acceleration of an electron hole occurring at its initiation because of interaction with an ion stream.

There are different ways to initialize an electron hole at the start of a numerical simulation. One popular way is to create an electron distribution and potential distribution that represents a self-consistent hole *in the absence of ion response*, that is, with uniform ion density. Schamel's electron distribution form is a natural choice. Then at time $t = 0$ the ion dynamics is turned on and the hole evolves. It is observed in simulations with physical mass ratios[16, 27, 19] that, after a short period of quiescence, the hole accelerates almost instantaneously to a fraction of the electron thermal speed, and moves away, leaving behind a small ion density perturbation. We call this "electron density deficit initialization". The ions do not experience the *growth* of the potential, since it is already present when their dynamics is turned on.

An alternative is to initialize a hole consisting of a local distortion of $f_e(v)$ formed by depleting the distribution of electrons at speeds near the desired hole speed and spreading them out across all other speeds (but at the same spatial location). The simulation is therefore initialized with zero electron and ion density perturbation, and so with uniform potential. On the electron transit time scale, the enhancement of the non-resonant part of the electron distribution moves away from the spatially localized region, leaving behind the electron deficit at the hole speed, which self-organizes into an electron hole. We call this "uniform density initialization". It gives a somewhat smoother startup, and exposes the electrons (and to some extent the ions) to the growth of the initial potential.

Simulations that turn off the ion response, regarding them as simply a uniform neutralizing background, give none of the transient dynamics we are discussing.

## 4.4 Electron density deficit initialization

The initially uniform ions, presumed to have essentially zero speed for an initially stationary hole, instantly experience the localized potential peak of the hole. They begin to accelerate away from the hole on either side. If the hole remains stationary (or if the hole breaks up into two symmetric parts that move in opposite directions[13]), then the net ion momentum change is zero by symmetry. However, if the hole acquires (perhaps by a random or imposed initial perturbation) a small velocity in one direction or the other, more ion momentum is expelled in the direction of the hole velocity. The net ion momentum then increases in the $U$-direction. Electrons must absorb some, at least, of that initial transient momentum. Since they are moving much faster than the ions, they can plausibly be treated in the short-transit-time approximation. The ions cannot consistently be treated using the short-transit-time approximation during the initial transient. However, it is worth noting that eq. (47) predicts growth of potential by a large factor will give a large factor increase in the ion speed in the hole frame. In other words it predicts initial instability. The electron momentum jetting rate of change (eq. 48) is in the $-\dot{U}$-direction. Therefore, since balancing the $+U$-direction ion momentum increase requires electrons to gain momentum in the $-U$-direction, the hole acceleration $\dot{U}$ must be in the $+U$-direction, the *same* direction as $U$. Evidently this is an unstable situation. The hole "accelerates itself" in whichever direction it begins to move.

We therefore expect there to be unstable increase of hole velocity on the electron timescale. What speed will the hole ultimately reach? Well, if the hole acceleration is practically instantaneous on the ion timescale, then the initially stationary ions become non-resonant so fast that we can ignore the jetting effect on the ions ($\dot{p}_o$) during the transient. Also, they experience negligible hole growth, ($\dot{p}_g$). If so, then the final hole speed is obtained by requiring the total electron and final contained ion momenta to cancel (in the frame of the external ions).

The final contained ion momentum density (change) is eq. (11)

$$\delta p = m(n - n_1)\delta U = mn_1 \delta U \left(\left|\frac{v_1}{v}\right| - 1\right) \simeq mn_1 U \left(\frac{\varphi_i}{mU^2}\right), \tag{59}$$

in which $v_1 = -\delta U = -U$, and $v = \pm\sqrt{U^2 - 2\varphi_i/m_i}$. The total electron momentum change is the time-integral of $\dot{p}_o + \dot{p}_c + \dot{p}_t + \dot{p}_g$ given in section 3.4. But if we ignore potential growth we might use the approximation eq. (48). Then the electron momentum time-integrates to $-n_e m_e U h(\chi)$ (recall $\chi^2 = |\varphi|/T_e$). Setting this equal to minus $\delta p$, we get

$$U^2 = \frac{\varphi_i}{m_e h} = \frac{T_e}{m_e}\frac{\chi^2}{h(\chi)}. \tag{60}$$

The final hole speed, even for shallow holes, for which $h \simeq \chi^2$, is thus approximately the electron thermal speed $U \sim \sqrt{T_e/m_e}$. This fact compromises the presumption in our calculation of $h$ that the electron Maxwellian distribution is unshifted relative to the hole. So the calculation cannot be expected to be precise. Nevertheless the published simulations[16, 27, 19], for deep holes $|\varphi|/T_e > 1$, observed final hole speed of $0.55\sqrt{T_e/m_e}$, which is qualitatively consistent with our estimate.

## 4.5 Uniform density initialization

Uniform density initialization is observed in simulations[26] to give rise to a slower and more moderate initial transient. Fig. 3 gives an example. Holes initialized symmetrically at zero speed (the ion mean speed) dwell for a time of perhaps $50/\omega_{pe}$ and then accelerate to typically $6c_s$ in a similar time. The dwell period is probably the time required for the initial instability to grow an asymmetric perturbation, because when initialized with $U_0 \neq 0$ (as in Fig. 3) the dwell period is practically absent. Also the increase $U - U_0$ during the transient is smaller, for example when $U_0 = 10c_s$, $U - U_0 \approx 2c_s$.

The duration of the acceleration period is short enough that the short transit-time approximation is barely adequate for slow ions. In a time of $50/\omega_{pe}$, ions at the sound speed travel a distance $50\sqrt{m_e/m_i}\lambda_{De} \sim \lambda_{De}$. Hole spatial extent is generally a few times $\lambda_{De}$ the Debye length. We find nevertheless[26] that the predictions of of Fig. 4 are in good agreement with simulations. Shallow holes grown from zero initial velocity reach no more than about $U/c_s \simeq (m_i/m_e)^{1/4} \simeq 6$, and holes growing from higher initial speeds accelerate less.

## 4.6 Hole trapping between two ion streams

One of the motives for the present study was the observation that in PIC-simulated wakes electron holes are spawned by a kinetic electron instability[18]. Fig. 5 illustrates the observations. Holes with speeds much greater than $c_s$ move out of the simulation along orbits that are practically the same as the background electrons, as determined by the large-scale potential gradients of the wake as a whole. That is in agreement with section 3.1. However, the most important holes are a different class that have almost zero velocity on the electron scale and therefore remain in the wake for a long time, growing by a new mechanism[28] until they cause nonlinear disruption of the ion streams. The key question is, what prevents those holes from acquiring substantial velocity and moving out of the wake?

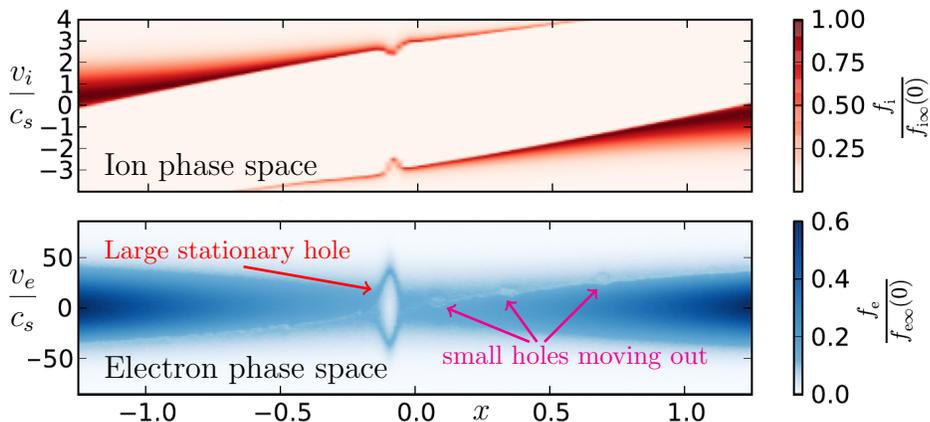

Figure 5: Distribution function contours in phase space of a wake simulation, showing small convecting holes and a large stationary hole. The $x$ position is in units of the object radius. After ref [18].

The present analysis provides the answer. In a wake there are two ion streams that enter the wake from opposite sides, shown in the ion phase space of Fig. 5. Their distributions aren't exactly delta functions in velocity, but they are fairly narrow, and they have zero distribution function for a significant velocity range between them. Near the center of the wake, the ion streams move in opposite directions at a few times the sound speed. If a hole has a velocity that lies *between* the velocities of the two ion streams, then its interaction with the ions of the streams, based upon the present analysis, acts to keep it there. The streams act as barriers in velocity space, that the hole cannot approach or cross. If the hole "attempts" to approach a stream, it is "pushed" back by the mechanism of section 4.1. The "pull" of the opposite stream becomes small as it moves away from it. The strength of the push far exceeds the strength of the background field. Consequently holes trapped between the ion speeds remain trapped. In fact the structure of the wake is such that the mid-point velocity between the streams varies with transverse position in the wake. The holes that grow to large size are observed to have speeds that vary with their transverse position consistent with their being trapped between the ion streams.

# 5 Summary

We have analyzed the momentum changes induced by growth or acceleration of a one-dimensional collisionless electron hole, using the short-transit-time approximation, including its interaction with ions. The momentum conservation equations derived are analytically solved in the shallow-hole approximation. The results are qualitatively and sometimes quantitatively in agreement with observations of numerical simulations. The analysis shows why holes move and accelerate in the way they do.

# Acknowledgements

This work was inspired by the simulations of Christian Haakonsen using the PIC code he developed. His early input and valuable discussions are gratefully acknowledged. The work was partially supported by the NSF/DOE Basic Plasma Science Partnership under grant DE-SC0010491.